\documentclass[12pt,preprint]{aastex}
\begin{document}

\parindent=1.0cm

\baselineskip=0.8cm

\title{THE OPEN CLUSTER NGC2437 (Messier 46) \altaffilmark{1}}

\author{T. J. Davidge}

\affil{Dominion Astrophysical Observatory,
\\National Research Council of Canada, 5071 West Saanich Road,
\\Victoria, BC Canada V9E 2E7}

\altaffiltext{1}{Based on observations obtained with MegaPrime/MegaCam, a joint
project of CFHT and CEA/DAPNIA, at the Canada-France-Hawaii Telescope (CFHT)
which is operated by the National Research Council (NRC) of Canada, the
Institut National des Sciences de l'Univers of the Centre National de la
Recherche Scientifique (CNRS) of France, and the University of Hawaii.}

\begin{abstract}

	The stellar content of the open cluster NGC 2437 (Messier 46) is investigated 
using moderately deep $u*,g',$ and $r'$ MegaCam images. 
When compared with solar metallicity isochrones, the
$(g', u'-g')$ and $(r', g'-r')$ CMDs are consistent with 
an age log(t$_{yr}) = 8.35 \pm 0.15$, a distance modulus $\mu_0 = 
11.05 \pm 0.05$, and a color excess $E(B-V) = 0.115 \pm 0.035$. The 
$r'$ luminosity function (LF) of main sequence stars in the magnitude range 
$r' < 17$ (i.e. masses $\gtrsim 0.8$ M$_{\odot}$) has a shape that follows 
solar neighborhood star counts. However, at fainter magnitudes the cluster 
LF is flat, in contrast with what would be expected from solar 
neighborhood counts. The clustering properties of stars in NGC 2437 
are investigated by examining the two-point angular correlation functions of 
main sequence stars in different brightness ranges. Main sequence stars fainter than 
$r = 17$ are less centrally concentrated than brighter stars and are found over a 
larger area of the sky, suggesting that there is a corona of faint main 
sequence stars around NGC 2437. Based on the flat LF and extended spatial 
distribution of faint stars, it is concluded that NGC 2437 is actively 
shedding stars with masses $\lesssim 0.8$ M$_{\odot}$.

\end{abstract}


\section{INTRODUCTION}

	Star clusters are basic targets for observational studies of stellar 
and galactic evolution. However, the majority of clusters are 
short-lived, being disrupted after only a few Myr (e.g. Lada \& Lada 
2003) probably due to large-scale mass loss driven 
by supernovae and stellar winds. The clusters that survive their infancy 
are likely those with initial conditions that (1) allow them 
to develop regions early-on where stars are the dominant mass component, 
thereby reducing the disruptive influence of gas loss, and/or (2) are in a contracting 
phase of their dynamical evolution at the time of peak gas expulsion (e.g. Smith 
et al. 2011). Clusters that are seen at the present day may then be biased examples 
of the majority of clusters that initially formed.

	The spatial distribution of stars within 
clusters provide insights into their past history. 
Clusters that are younger than a few tens of Myr can be surrounded by 
diffuse halos that contain some of the most massive cluster members 
(e.g. Davidge 2012). The spatially dispersed nature of these massive stars is likely 
the consequence of the large-scale loss of gas early in 
the evolution of the cluster (Bastian \& Goodwin 2006). The central bodies of older 
clusters are surrounded by coronae populated by low mass stars (e.g. Kholopov 
1969), which are expected to move to the cluster periphery due to 
dynamical relaxation (e.g. Prisinzano et al. 2003; Sharma et al. 2006). Stars in 
the outermost cluster regions will be the most susceptible to stripping, and as 
clusters in the disk disperse they may leave debris trails that extend over large 
swaths of the sky (e.g. Chumak \& Rastorguev 2006; Chumak et al. 2009). 

	In the present study, $u*, g',$ and $r'$ images obtained 
with the Canada-France-Hawaii Telescope (CFHT) MegaCam are 
used to examine the stellar content of the open cluster NGC 2437. NGC 2437 is an 
interesting target for investigation as it is nearby and has a mass and age 
that make it pertinent for studies of cluster disruption. Basic 
properties of NGC 2437, taken from the WEDBA database (Mermilliod 1995), are 
listed in Table 1. 

	It is somewhat surprising that -- despite being a relatively nearby and 
well-populated intermediate age cluster -- NGC 2437 has not 
yet been extensively studied with electronic detectors. Sharma 
et al. (2006) use CCD images to investigate stars in NGC 2437 to as 
faint as $V \sim 18 - 19$, and find a distance of 1.51 kpc ($\mu = 10.9$) and an age 
log(t$_{yr}) = 8.4$. Their data trace the radial distribution of stars out to 20 
arcmin from the cluster center. Sharma et al. (2006) find that NGC 2437 is more spatially 
extended in the near-infrared than at visible wavelengths, hinting that 
the redder (lower mass) cluster members may be more widely distributed than the bluer 
(higher mass) members. Still, despite this hint that mass segregation may be underway, 
Sharma et al. (2008) find that stars as faint as $V \sim 18 - 19$ in NGC 2437 have a 
mass function exponent that is consistent with the Salpeter (1955) value. 

	The specific goals of the present study are (1) to estimate the age, reddening, 
and distance of NGC 2437, and (2) to search for signatures of cluster dissruption by 
examining the cluster luminosity function (LF) and the spatial distribution of cluster 
stars spanning a range of masses. A discussion of the observations and their reduction is 
provided in \S 2. Particulars of the photometric measurements are discussed in 
\S 3, while the cluster CMDs and LFs are examined in \S\S 4 and 5. The 
spatial distribution of stars in the cluster is investigated in \S 6. 
A summary and discussion of the results follows in \S 7. The use 
of different criteria for identifying objects with suspect photometry and the 
possible relationship between NGC 2437 and the planetary nebula (PN) NGC 2438 are 
investigated in the Appendix.

\section{OBSERVATIONS}

	The data were recorded with the CFHT MegaCam (Boulade et al. 2003) 
as part of program 2010BD89. The detector in MegaCam 
is a mosaic of thirty-six $2048 \times 4612$ E2V CCDs arrayed in a 
$4 \times 9$ format with 0.185 arcsec pixel$^{-1}$ sampling. There are 
80 arcsec gaps between the three CCD banks, and 13 arsec gaps between individual CCDs.
A single exposure images a $\sim 1$ degree$^2$ field.

	Details of the observations are summarized in Table 2. Images 
with short and long integration times were recorded to extend the 
magnitude range of detected stars. The minimum exposure time in 
$u*$ was longer than in the other filters to avoid potential charge transfer problems 
that could arise from the low sky levels recorded through this filter during dark time.

	Initial processing was done with the CFHT ELIXIR pipeline, and this included 
bias subtraction and flat-fielding. The ELIXIR-processed images were 
corrected for positional offsets between images that are recorded in different 
filters. These offsets vary in size with location in the science field, and are 
typically a few pixels (i.e. $< 1$ arcsec) near the edge of the MegaCam field when 
$u*$ and $g'$ images are aligned near the center of the detector mosaic. Offsets 
between the $g'$ and $r'$ images are much smaller. 
These distortions were corrected in a differential manner by 
using the IRAF GEOMAP and GEOTRANS routines to map the $u*$ and $r'$ images of 
each cluster into the reference frame defined by the $g'$ image. 

\section{PHOTOMETRIC MEASUREMENTS}

\subsection{Stellar Brightnesses}

	Target lists and point spread functions (PSFs) 
were obtained by running routines in the DAOPHOT (Stetson 1987) package. 
There are many unsaturated stars with high S/N ratios in the images, 
and PSFs were constructed by combining 50 -- 100 of the most isolated of these 
in each filter. Faint companions were removed iteratively by subtracting 
progressively improved versions of the PSF.

	Stellar brightnesses were measured with the 
PSF -- fitting routine ALLSTAR (Stetson \& Harris 1988). Sources that are not 
well-matched by the PSF, and/or that hinder convergence to a stable solution are 
rejected by ALLSTAR. Still, the photometry catalogue that is output by ALLSTAR 
contains sources with poor photometry, and some of these 
can be identified using information that is provided by ALLSTAR. Following 
Davidge (2010), such objects in the CFHT images were identified using 
$\epsilon$, the error in magnitude that is computed by ALLSTAR. 

	Stars that are neither saturated nor blended define a 
monotonic relation between $\epsilon$ and magnitude. Objects that 
depart from this relation tend to be background galaxies (e.g. Figure 2 of 
Davidge 2010), unresolved blends, saturated stars, or cosmetic defects, 
and these were removed from the photometric catalogues. Objects with $\epsilon 
\geq 0.3$ magnitudes, the majority of which are near the faint limit 
of the photometry, were also deleted. The use of other ALLSTAR-computed indices to 
identify objects with suspect photometry is investigated in the Appendix. 

\subsection{Calibration}

	Instrumental magnitudes were transformed into 
the SDSS system by applying linear transformation relations with 
zeropoints obtained from observations of photometric standards that are recorded 
during each MegaCam observing period. The $u* \rightarrow u'$ transformation 
warrants some discussion, as the effective wavelengths of the $u*$ and $u'$ filters 
differ by $\sim 200\ \AA$, in the sense that $u*$ is redder. This difference in 
effective wavelength is particularly acute as the $u'$ filter samples the Balmer 
discontinuity. The difference between the filter passband then affects the ability 
to obtain reliable $u'$ magnitudes for intermediate spectral-type 
high surface gravity stars from $u*$ magnitudes (Clem et al. 2008). While previous 
experience demonstrates that a reliable $u* \rightarrow u'$ 
transformation can be acheived using a linear transformation equation, that result is 
restricted to intrinsically luminous stars with weak Balmer breaks (e.g. 
Davidge \& Puzia 2011; Davidge et al. 2012).

	Bright main sequence stars in NGC 2437 have an intermediate 
spectral-type, and so have strong Balmer discontinuities. 
Because cluster members have a (more-or-less) common, measureable reddening, their 
$u_*$ magnitudes could be transformed into $u'$ magnitudes using 
-- say -- a non-linear relation like that found by Clem et al. 
(2008), which would result in corrections of up to $\sim 0.4$ magnitudes 
being applied to the $u*$ magnitudes of late B and early A-type main sequence stars. 
However, the application of a non-linear relation that involves large corrections 
to field stars, which have a range of reddenings and evolutionary 
states, is much less secure, as there is the potential 
to produce final magnitudes that are greatly in error. For example, the 
large transformation corrections from Clem et al. (2008) that are appropriate 
for A and B main sequence stars would be applied in error to stars of later 
spectral-type that have had their reddening underestimated. To prevent introducing 
such large errors into the transformed magnitudes, a psuedo $u'$ magnitude, 
$u'_*$ was computed for each source using a linear transformation relation. In \S 4 
it is demonstrated that the SEDs of stars with $u'_*$ magnitudes match those 
computed using $u'$ over most of the color range sampled in NGC 2437.

\subsection{Artificial Star Experiments}

	Sample completeness and uncertainties in the photometry 
-- both random and systematic -- were assessed by running artificial star experiments. 
The artificial stars were assigned colors and brightnesses 
that are representative of objects in NGC 2437. As with the actual 
observations, an artificial star was considered to be detected only if it was 
recovered in at least two filters (either $u' + g'$ or $g' + r'$). The photometric 
catalogues produced by ALLSTAR as part of these experiments were filtered with 
the $\epsilon$-based criteria that were applied to the science data. 

	Errors in the photometry were estimated by comparing the 
input and measured brightnesses of artificial stars, and then computing the 
mean difference and the dispersion about the mean at various magnitudes. 
Completeness and the uncertainties in the photometry are coupled, in the sense 
that random and systematic uncertainties in the photometry climb 
rapidly towards fainter magnitudes when the completeness is below 
50\%. The magnitude at which 50\% completeness occurs, which is 
near $u' = g' = r' = 23$ in the deep exposures of NGC 2437, is thus one 
measure of the photometric faint limit. 

\section{RESULTS: CMDs}

	The $(g', u'_*-g')$ and $(r', g'-r')$ CMDs of sources in and around NGC 2437 
are shown in Figures 1 and 2. Measurements with $g' < 14.5$ and 
$r' < 14.5$ were obtained from the short exposure MegaCam images. The open squares are 
stars with $V < 10.5$ from Lynga (1960), and the published magnitudes were transformed 
into the SDSS system using the relations from Smith et al. (2002) and Jordi et al. 
(2006). The envelope of $\pm 2\sigma$ random errors computed from the artificial 
star experiments is shown in the lower left hand corner of each CMD.
The uncertainties in the photometry become substantial for 
objects fainter than $g' = r' = 20$. 

	The main sequence of NGC 2437 cuts diagonally across both CMDs, although the 
ability to trace it is compromised at some points by contamination from non-cluster 
objects. These contaminants are a mix of field stars and galaxies, although the 
contribution from galaxies should be modest given the low Galactic latitude of NGC 2437. 
In addition to producing the diffuse spray of points that is scattered over 
both CMDs, field stars populate two distinct features, the 
most prominent of which is a near-vertical blue plume. 
This feature intersects the main sequence of NGC 2437 near $g' \approx r' \approx 15$, 
complicating efforts to trace the cluster sequence in this part of the CMD. Field 
main sequence stars also populate the vertical plume near the faint end of the $(r', 
g'-r')$ CMD. This feature intersects the faint end of the cluster main sequence.

	The $(u'_*-g', g'-r')$ two-color diagram (TCD) is shown in Figure 3. The cyan 
line is the locus of main sequence stars traced from the $(g', u'_*-g')$ and 
$(g', g'-r')$ CMDs of NGC 2437. The red line is the locus of Galactic 
main sequence stars from Johnson (1966), transformed into the SDSS system 
using relations from Smith et al. (2002) and Jordi et al. (2006), and then 
reddened to match the $E(B-V)$ of NGC 2437 using the relations in Table 6 
of Schlegel et al. (1998). The agreement between these two sequences 
suggests that the $u'_*$ magnitudes are reasonable proxies for SDSS $u'$ magnitudes 
over much of the color range sampled by these data.

	The spectral-energy distributions (SEDs) of many 
non-cluster objects will differ from those of cluster sources, raising the prospect 
that at least some contaminating objects can be identified based on their location 
on the TCD. Consider the ridgeline of the field star plume, which is shown as 
the cyan line in Figure 3. This is an almost vertical feature on the CMDs, 
and so occupies a comparatively modest range of colors in the TCD when compared with 
solar neighborhood and NGC 2437 main sequence stars. Still, 
field stars on this sequence are offset $\sim 0.1$ magnitude along the $(g'-r')$ axis 
from the cluster sequence on the TCD -- at least some of 
the sources on this very prominent feature can be identified based on their 
locations on the TCD.

	For the current study, two samples of objects are defined based on their 
location on the TCD. `Cluster' stars are those that lie within 
$\pm 0.275$ magnitude in $u'_*-g'$ of the NGC 2437 main sequence on the TCD, while 
`non-cluster' objects are those that occupy the rest of the TCD. Experiments 
indicated that this color envelope is the narrowest that preserves cluster 
main sequence stars. The application of a tighter color envelope about the NGC 2437 
sequence on the TCD will reject more non-cluster objects, but at the expense of 
also rejecting cluster main sequence stars.

	The CMDs of objects in the cluster and non-cluster samples are compared in 
Figures 4 and 5, where the data have been sorted into three radial intervals that sample 
equal areas on the sky to facilitate an assessment of residual contamination 
from non-cluster objects. While there is residual non-cluster contamination in 
the cluster sample, the contamination in the upper rows of Figures 
4 and 5 is substantially smaller than in Figures 1 and 2. 
There is no obvious binary sequence above the cluster main sequence, suggesting that 
the fraction of binaries with mass ratio near unity is modest in NGC 2437. 

	It is evident from Figures 4 and 5 that all but the bluest field plume 
stars have been identified as non-cluster objects. The majority of stars 
that occupy the plume with $r' > 18$ and $g'-r' \sim 1.5$ in the $(r', 
g'-r')$ CMD are also identified as being non-cluster in origin. 
The number of non-cluster objects in each radial interval is more-or-less 
constant, as expected if these objects are uniformly distributed on the sky. 

	Isochrones with $Z = 0.019$ from Marigo et al. (2008) are compared 
with the cleaned inner annulus CMDs in Figure 6. The distance modulus 
and reddening listed in the WEDBA database as of mid-2012 have been assumed 
for this initial comparison. As discussed in \S 3, the MegaCam data are not in the 
SDSS $u'$ system. Therefore, the isochrones were transformed 
into the $u'_*$ system by finding the offset in $u'-g'$ between 
the observed and model main sequences at a given $g'-r'$ color on 
the $(u'-g',g'-r')$ TCD. This offset was then applied to the model $u'$ magnitudes. 

	The isochrones in Figure 6 fall above the cluster main sequence, 
indicating that the adopted distance modulus is too small. 
The distance modulus and reddening were adjusted to obtain better agreement 
with the observations, and the results for $\mu_0 = 11.05$ and $E(B-V) = 0.115$, 
which are adopted for the remainder of the paper, are shown in Figure 7. Comparisons 
with a grid of $\mu_0$ and $E(B-V)$ values indicated that variations in these quantities 
that exceed $\pm 0.05$ magnitude ($\mu_0$) and $\pm 0.035$ magnitude ($E(B-V)$) 
could be identified, and so these are adopted as the estimated uncertainties.
The reader should keep in mind that these uncertainties assume that the stars 
in NGC 2437 have a known metallicity (Z = 0.019).

	The models shown in Figure 7 suggest an age log(t$_{yr}) = 8.5$ based 
on the $(g', u'_* - g')$ CMD, and log(t$_{yr}) = 8.2$ based on the 
$(r', g'-r')$ CMD. Therefore, log(t$_{yr}) = 8.35 \pm 0.15$
is adopted for NGC 2437, and additional support for this comes from 
the bright end of the cluster LF (\S 5). Isochrones with this age do not pass 
through the two bright blue objects observed by Lynga (1960). To the extent that the 
models are correct, then this suggests that these stars 
(1) are not cluster members, (2) are variable stars, (3) are binaries, and/or (4) 
are not experiencing `normal' evolution of the type assumed by the models.

\section{RESULTS: LFs}

	The $r'$ LF of objects that (1) have $g'-r'$ within $\pm 0.15$ magnitudes 
of the NGC 2437 main sequence on the $(r', g'-r')$ CMD, and 
(2) are within 19.3 arcmin of the center of NGC 2437 -- 
where the density of cluster stars is highest (Figures 4 and 5) -- is shown in Figure 8. 
The faint limit of the LF is defined by the $g'$ measurements, 
and the artificial star experiments indicate that the data are 100\% complete for 
$r' < 20$, which corresponds to $\gtrsim 0.5$ M$_{\odot}$. The LF in Figure 8 
has been corrected for non-cluster objects using source counts in the outer annulus.

	The $r'$ LF can be divided into three magnitude regimes. At the bright end 
($r' < 13$), the LF climbs steadily towards fainter magnitudes. There is then a break 
in the LF at $r' \sim 13$, which is due to an inflexion in the mass--magnitude 
relation near $\sim 2$ M$_{\odot}$. Finally, the LF is more-or-less flat when 
$r' > 16$ (i.e. for masses $\leq 1$ M$_{\odot}$). It is demonstrated in the Appendix 
that the flat nature of the LF at these magnitudes is a robust result that is 
reproduced using different criteria to cull sources with poor photometric measurements.

	Model LFs, generated from the Marigo et al. (2008) isochrones and downloaded from 
the {\it Padova database of evolutionary tracks and isochrones} web 
site, \footnote[2]{http://stev.oapd.inaf.it/cgi-bin/cmd} are also shown in Figure 8. 
The models assume Z = 0.019, and have ages $t = 200$ Myr (dashed line) 
and 300 Myr (dotted line). A Kroupa (1998) Initial Mass 
Function (IMF), which is based on solar neighborhood star counts, 
was adopted. The models have been scaled to match the LF between $r' = 13.5$ and 15.5.

	The change in the character of the LF near $r' \sim 13$ provides a loose check 
on the distance modulus, and the agreement with the models at this magnitude suggests 
that the adopted distance modulus is correct to within $\pm 0.3$ magnitude. While the 300 
Myr model provides a better match to the observations at the bright end 
than the 200 Myr model, there is a tendency for both models to fall above the NGC 2437 
LF at the bright end. Near $r' = 12.5$ the oberved number counts differ from the 300 Myr 
models at more than the $2\sigma$ level. Better agreement with the model LFs would 
result with an apparent distance modulus that is $\sim 0.1$ dex greater than 
deduced from the CMDs, although this would come at the expense of degrading the 
agreement between the isochrones and the cluster main sequence on the CMDs at 
intermediate brightnesses.

	The models depart significantly from the observed LF when $r' > 17$, 
with the size of the disagreement increasing steadily towards fainter $r'$. At $r' = 20$ 
(main sequence mass $\sim 0.5$ M$_{\odot}$), the number of stars predicted by the 
model is roughly an order of magnitude greater than observed. 
The LF of the inner regions of NGC 2437 thus differs significantly from that of the 
solar neighborhood when $r' > 17$ (i.e.  $\leq 0.8$ M$_{\odot}$). Possible causes of 
this difference are examined in \S 7. 

\section{CLUSTER STRUCTURE}

	The spatial distribution of stars in a cluster contains clues about 
its past history. The spatial distribution and clustering properties 
of stars in and around NGC 2437 are examined in this section using two 
approaches. The two-point angular correlation function (TPCF) of cluster 
stars is examined in \S 6.1, while the physical distribution of cluster stars 
on the sky is considered in \S 6.2. 

	As with the analysis of the LF in \S 5, 
objects within $\pm 0.15$ magnitude in $g'-r'$ about the main sequence ridgeline 
are adopted as cluster members, although the result of applying a narrower 
extraction window is also investigated. Cluster stars are divided into various brightness 
regimes using the comparison between the observed LF and models in Figure 8 as a guide.
The BMS (bright main sequence) sample includes the most massive 
main sequence stars in the cluster, which have $r' \leq 13$. The 
IMS (intermediate main sequence) sample includes objects from $r' = 13$ to 
$r' = 17$. The LF increases slowly towards fainter magnitudes in this magnitude 
interval, at a rate that is more-or-less consistent with that expected from the 
models shown in Figure 8. The LMS (lower main sequence) sample runs from 
$r' = 17$ to $r' = 20$. This is the magnitude interval where significant departures 
from the model LFs occur. Finally, a field star (FS) sample is defined that
consists of all objects that are outside of the $\pm 0.15$ magnitude 
main sequence extraction region. The objects in the FS sample are uniformly distributed 
across the field, and are used to remove signatures in the TPCF 
that arise from detector geometry -- such as gaps between detectors and the finite 
size of the MegaCam science field. 

\subsection{The TPCF}

	The TPCF multiplexes information by considering all possible object-object 
pairings in a sample, and so makes efficient use of the available information. However, 
as the correlation properties of all objects are considered in concert then 
there is not a direct relation between features in the TPCF and location within the 
area examined. Still, some generalizations can be drawn about the 
locations of groups that contribute signal to the TPCF. Basic 
geometric arguments imply that signal at the largest possible 
separations must originate near the edges of the sampled area.
The potential for ambiguity grows as progressively smaller separations 
are considered, as these separations can be measured over a larger fraction of the 
field. Another mitigating factor is the density distribution of stars in a cluster.

	The distributions of separations obtained for all possible object pairings 
(hereafter `separation functions' -- SFs) were calculated for sources in each 
of the samples. TPCFs were then constructed by dividing the SFs of 
the BMS, IMS, and FMS samples by the SF of the FS sample, and 
scaling the results according to the number of pairings.
The resulting TPCFs are compared in the top panel of Figure 9.

	The objects in the three samples have different clustering 
properties. The TPCF of the BMS sample has by far the largest amplitude, with power 
exceeding zero detected out to separations of $\sim 60$ arcmin. 
An inflexion point in the TPCF of the BMS sample occurs 
near separations of 20 -- 25 arcmin ($\sim 9 - 12$ parcsecs), 
hinting at a possible change in the structural characteristics of the cluster 
10 -- 12.5 arcmin from its center. A discontinuity is also seen in the TPCFs of 
IMS and FMS stars at this separation. 

	A complication in interpreting the TPCFs is that field stars are located over 
a wide range of distances along the line of sight, and so binaries with a range of 
different physical separations are included in each angular interval. 
Consider very wide stellar binaries, which have spatial 
separations $\sim 0.04$ parcsecs (e.g. Larson 1995; Simon 
1997). Binaries in the field with separations of 0.04 parsecs 
would be separated by a few arcmin if viewed at a distance of 100 -- 200 parcsecs, 
but similar binaries at the distance of NGC 2437 will 
have separations $\leq 5$ arcsec. It is thus of interest to consider the ratios of 
SFs of objects that are in the cluster.

	The ratio of the IMS and FMS SFs is shown 
in the lower panel of Figure 9. The IMS and FMS samples have similar clustering 
properties at separations $< 25$ arcmin. However, at larger separations IMS stars 
are less clustered. This is consistent with 
objects in the FMS sample contributing a larger fraction 
of the objects in the outermost regions of NGC 2437 than stars in the IMS sample. 

	The TPCFs of the IMS and FMS samples have substantially smaller 
amplitudes than the BMS TPCF. This reduction in amplitude is 
due -- at least in part -- to differences in contamination from 
field stars, which are uniformly distributed across the field and cause 
the TPCFs of field star-contaminated samples to be flatter 
than those of samples that are free of such 
contamination. Star counts in the inner and outer annuli (\S 5) suggest that as many 
as $\leq 60\%$ of the stars in the IMS and FMS samples may be field objects, 
while only $\leq 10\%$ of the BMS sample may be field stars. 

	The extent to which non-cluster objects affect the TPCFs 
can be assessed by comparing the TPCFs of samples 
constructed using different extraction widths along the color axis on the CMD. 
The fractional contamination from field stars is reduced as progressively narrower 
extraction windows are applied, although this reduction 
is obtained at the expense of also culling cluster stars that scatter 
about the main sequence on the CMD. To investigate the impact of non-cluster 
contamination on these data, a second set of TPCFs were computed for the IMS and FMS 
samples using a $\pm 0.05$ magnitude wide band in $g'-r'$. 
While not shown in Figure 9, the TPCFs of the IMS and FMS samples constructed 
with this narrower window have amplitudes that are $\sim 2 \times$ larger 
than those obtained with a $\pm 0.15$ magnitude cut. This is consistent with reduced 
levels of non-cluster contamination. 

	The ratio of the SFs constructed with the $\pm 0.05$ mag extraction window is 
shown as a dashed line in the lower panel of Figure 9. The ratio of the IMS and FMS 
SFs obtained with the $\pm 0.05$ mag window differs from that obtained with 
the $\pm 0.15$ mag window, in the sense that the IMS sample is more centrally 
concentrated than the FMS sample when the narrower extraction window is used. 
This comparison reinforces the result discussed above that the cluster
stars in the FMS sample are more diffusely distributed than those in the IMS sample.

\subsection{The Projected Distribution of Stars}

	Additional insights into the structure of NGC 2437 
can be obtained by visually examining the spatial 
distribution of objects. The distributions of sources in 
the BMS, IMS, and FMS samples are shown in Figure 10. The top row shows the 
source distributions of objects that fall within a $\pm 0.15$ mag extraction window 
about the main sequence locus on the $(r', g'-r')$ CMD, while the bottom rows shows 
the distribution of sources within a $\pm 0.05$ mag extraction window. The pixel 
intensities reflect the number of stars within $1.5 \times 1.5$ arcmin gathers. This 
coarse bin size was adopted to suppress stochastic fluctuations in the star counts. 
still, the BMS distributions are dominated by single objects outside the main 
body of the cluster.

	Features that are present in the top row of Figure 10 
also tend to be present in the bottom row. That 
the spatial distributions of each sample type are not sensitive to the width of the 
extraction window on the CMD suggests that prominent structures in Figure 10 
are dominated by cluster stars, and not field stars. The 
objects in the various samples have different spatial 
distributions, with the BMS and IMS samples being more compactly distributed than the 
FMS sample. This is consistent with the comparisons between the TPCFs of the samples 
in the previous section.

	The IMS and FMS distributions have irregular morphologies at large radii, and 
the FMS distributions show a concentration of objects to the 
south and west of the cluster center. This feature is likely not due to 
field stars, as it is equally strong in the top and bottom rows of Figure 10, 
indicating that it is populated by objects with photometric properties that tightly 
follow the cluster main sequence. The nature of this feature is discussed further 
in the next Section.

\section{SUMMARY AND DISCUSSION}

	The stellar content of the open cluster NGC 2437 
(Messier 46) has been investigated. Despite being a moderately massive nearby 
intermediate age cluster, NGC 2437 has not yet been extensively surveyed with 
modern detectors; the one exception is the study by Sharma et al. (2006). 
The angular size of NGC 2437 is well-suited to the 
1 degree$^2$ MegaCam science field, and the data discussed here are the 
deepest wide-field images of this cluster that have been recorded to date. 

	The $(g', u'-g')$ and $(r', g'-r')$ CMDs of NGC 2437 have been compared 
with solar metallicity isochrones, and a distance modulus $\mu_0 = 11.05 \pm 
0.05$ and a color excess $E(B-V) = 0.115 \pm 0.035$ were obtained. 
These comparisons also suggest an age $220^{+100}_{-60}$ years. Star counts near the 
bright end of the $r'$ LF favor values near the upper limit of this age range. 
The distance modulus, reddening, and age deduced for NGC 2437 from the MegaCam 
data are not significantly different from those found by Sharma et al. (2006). 
However, the distance modulus found here is larger than that computed by Stetson (1981) 
from multicolor photoelectric observations of A stars. 
Given its age, NGC 2437 likely formed as part of the 
recent `local starburst' (e.g. Bonatto \& Bica 2011).

	The main sequence of NGC 2437 can be traced to $r' > 20$ in these data, 
allowing the LF of stars with masses $\geq 0.5$ M$_{\odot}$ 
to be investigated. After correcting for contamination by non-cluster objects, 
the shape of the LF for stars with $r' < 17$ ($\geq 0.7$ M$_{\odot}$) 
within 19.3 arcmin of the cluster center is consistent with 
solar neighborhood star counts. However, at fainter magnitudes the central regions 
of NGC 2437 appear to be deficient in stars.

	At first glance, such a deficiency among faint stars seems to contradict 
the results from Sharma et al. (2008), who find that the mass function of NGC 2437 has an 
exponent that (1) is comparable to the Salpeter (1955) value, and (2) is consistent 
with that of other clusters in their sample. However, the faint limit of the 
Sharma et al. (2008) data is $V \sim 18$, which corresponds to $r' \sim 17$. Their data 
is thus limited to the magnitude range that the MegaCam data finds is in agreement with 
the models.

	Velocity dispersion measurements suggest that there is 
a high binary fraction in NGC 2437 (Kiss et al. 2008), while modelling of the cluster CMD 
suggests that the binary fraction may be 30 -- 40\% (Sharma et al. 2008). If stars in 
binary systems form in numbers that follow the IMF, then binarity will have the 
greatest impact near the faint end of the LF, where the incidence of binaries with mass 
ratios approaching unity will be highest. While the model LFs considered here do not 
include binaries, a deficiency of stars with sub-solar masses would still 
remain in NGC 2437 if models that include binaries had been used. Indeed, when 
compared with models that do not include binaries the addition of binaries
typically changes the mass function exponent inferred from LFs by only a few 
tenths (e.g. Sagar \& Richtler 1991; Sharma et al. 2008). 
This is roughly an order of magnitude smaller than what is needed to explain the 
differences between the models and observations at faint magnitudes in Figure 8. 

	A deficiency in stars at the faint end of the 
NGC 2437 LF could occur if stars with masses $\leq 0.7$ M$_{\odot}$ 
have been dispersed from the central regions of the cluster. This 
could result from mass segregation, the structural signatures of which may be 
detectable after only a few cluster crossing times. Prisinzano et al. (2003) 
find evidence for mass segregation in the $\sim 100$ Myr cluster NGC 2422, while 
Delgado et al. (2011) find evidence of mass segregation in even younger clusters. 
Sharma et al (2006) find differences in open cluster radii measured from 
optical and 2MASS data, in the sense that some clusters -- 
including NGC 2437 -- are more extended spatially in the near-infrared than at shorter 
wavelengths. Given that the optical and near-infrared wavelength regimes are 
sensitive to different stellar mass ranges, then this suggests that the 
structural properties of some clusters in their sample, including NGC 2437, 
depend on stellar mass in a way that is consistent with mass segregation. The 
investigation of the TPCF in \S 6 provides further evidence to support 
the presence of mass segregation, as it indicates that the clustering properties of stars 
in NGC 2437 depend on magnitude, in the sense that bright main 
sequence stars have a more compact spatial distribution than faint main sequence stars. 

	A picture is thus emerging in which NGC 2437 has a diffuse corona that contains a 
larger fraction of stars with sub-solar masses than near the cluster center.
Such a diffuse circumcluster structure will be more susceptible to tidal 
stripping than the dense inner cluster environment, and so will likely be 
the source of stars that are shed from the cluster. 
Given that stars are fed into the peripheral regions of the cluster 
as mass segregation proceeds, then the corona will be replenished as 
stars are stripped away. In fact, the prognosis for the long-term survival of NGC 2437 is 
not good, and NGC 2437 may actively be shedding stars. Lamers et al. (2005) find that 
a cluster in the Galactic disk with a mass 10$^4$ M$_{\odot}$ -- which is 
substantially larger than NGC 2437 -- may survive for only 1.3 Gyr. 
With a mass of $\sim 10^3$ M$_{\odot}$ (Sharma et al. 2008), then NGC 2437 will 
probably evaporate on the timescale of only a few disk crossing times. 

	Is there evidence of mass loss from NGC 2437? The spatial distribution of main 
sequence stars in Figure 10 indicates that low mass stars at large radii in 
NGC 2437 may extend to the south and west of the cluster, and we suggest that this 
may be the roots of a tidal debris trail. Such trails may not be rare in the 
Galactic disk, and simulations suggest that clusters that are a few 100 Myr 
older than NGC 2437 may leave debris trails that extent over many hundreds of parsecs 
(Chumak \& Rastorguev 2006). If the stars to the south and west of NGC 2437 are part 
of such a debris trail then they may belong to a much larger coherent structure, only a 
modest fraction of which can be captured with a single MegaCam pointing. 

\clearpage

\appendix

\section{AN INVESTIGATION OF DIFFERENT CULLING CRITERIA ON THE CLUSTER LF}

	A procedure for identifying and deleting objects with suspect photometry using 
$\epsilon$ is described in \S 3, and the dataset that was culled in this manner 
has been adopted as the baseline for this study. $\epsilon$ has 
the merit of being easily related to stellar magnitudes, 
and so provides -- at least in principle -- an intuitive basis for understanding 
biases that may be introduced by source culling. Still, $\epsilon$ is not the 
only parameter that can be used to identify objects with suspect photometric 
measurements. Here, the use of two other parameters that are computed by ALLSTAR -- 
sharp and $\chi$ -- to identify objects with suspect photometry are also explored. 
These parameters are described in detail by Stetson et al. (2003). They measure source 
characteristics that differ from those used to compute $\epsilon$, and so 
provide a means of identifying biases that may be introduced in the culling process.

	As is the case with $\epsilon$, sources with sharp and $\chi$ that depart 
significantly from the mean value at a given magnitude may be galaxies, stellar blends, 
or cosmetic artifacts. Following the procedure employed to filter the photometric 
catalogue using $\epsilon$, a function was fit to the 
envelope of the main body of objects on the sharp $vs.$ magnitude and $\chi \ vs.$ 
magnitude relations. Objects that fall outside of the fitted relations were then deleted. 
Near the faint limit of the data sharp and $\chi$ also show 
markedly greater scatter than at brighter magnitudes, and 
it was also decided to delete all objects with $|sharp| > 0.8$ and $\chi > 1.5$.

	The background-subtracted $r'$ LFs that result from cleaning the dataset 
using each parameter independently are compared in Figure A1. The faint end of the LF 
is shown in this figure, as this is where the majority of points are removed. Given that 
$\epsilon$, sharp, and $\chi$ characterize different source properties, it is 
not surprising that there are LF-to-LF differences. However, these 
differences tend to be modest in size. Indeed, the LF-to-LF 
dispersion falls within the uncertainties predicted by the number counts for the LF 
constructed using $\epsilon$ as the culling parameter. The flat nature of the $r'$ 
LF at the faint end is present in all three LFs, indicating that the difference between 
the model and the NGC 2437 LFs at the faint end that are 
discussed in \S 5 is not an artifact of the technique used 
to excise sources with suspect photometry.

\section{THE PLANETARY NEBULA NGC 2438}

	The PN NGC 2438 is located near the central regions of 
NGC 2437 on the sky. If NGC 2438 is a member of NGC 2437 then this makes that 
cluster of considerable importance for studies of the evolution of intermediate 
mass stars and PN. However, the radial velocity of NGC 2438 differs significantly from 
that of NGC 2437, suggesting either no physical connection or that the PN progenitor was 
ejected from the cluster (Kiss et al. 2008). 

	If NGC 2438 is associated with NGC 2437 
then the progenitor likely had a mass that is comparable to those of stars in 
the BMS sample. When compared with lower mass stars, the BMS stars are tightly clustered 
on the sky (\S 6), with a typical separation of $\sim 20$ arcmin (Figure 9), which 
corresponds to $\sim 10$ parsecs at the distance of NGC 2437. Assuming virial 
eqiuilibrium, then with a total cluster mass of $10^3$ M$_{\odot}$ the expected velocity 
dispersion is $\sim 1 - 2$ km sec$^{-1}$. For comparison, Kiss 
et al. (2008) find that the difference in velocity between NGC 2438 
and the cluster is 30 km sec$^{-1}$. If the progenitor of NGC 2438 formed as a 
member of NGC 2437 then its high velocity (and expected large physical displacement 
from the cluster center) makes it an extremely peculiar object when compared with 
other cluster members of similar mass.

	If NGC 2438 is not associated with NGC 2437 but is part of another -- 
not yet dissolved -- cluster that is viewed against/through NGC 2437 then a localized 
concentration of stars may be present near NGC 2438. There are two peaks in 
the distribution of FMS stars in the lower panel of Figure 10, and 
the peak immediately to the north of the MegaCam field center coincides 
roughly with the location of NGC 2438. In contrast, the BMS and IMS samples are more 
uniformly distributed near NGC 2438. Radial velocity and proper motion studies of 
objects within 5 -- 10 arcmin of NGC 2438 (the approximate extent of the peak in the 
FMS sample) should reveal if there is an ensemble of stars that is viewed fortuitously 
through/against NGC 2437. Of course, if the progenitor of NGC 2438 is older than a few 
hundred Myr then its host cluster may have long since dissolved, although in that case 
the PN might be part of an extended debris trail.

\acknowledgements{It is a pleasure to thank the anonymous referee for providing a 
timely and thoughful report that improved the manuscript. Sincere thanks are also 
extended to Dr. Christian Veillet who - as the director of the CFHT - granted the 
discretionary time for these observations.}

\parindent=0.0cm

\clearpage

\begin{table*}
\begin{center}
\begin{tabular}{ccccc}
\tableline\tableline
Name & E($B-V$) & $\mu$\tablenotemark{a} & log(t$_{yr}$) & [M/H] \\
\tableline
NGC 2437 & 0.154 & 11.17 & 8.4 & $+0.05$ \\
\tableline
\end{tabular}
\end{center}
\tablenotetext{a}{Distance modulus}
\caption{NGC 2437 Parameters from the WEDBA Database}
\end{table*}

\clearpage

\begin{table*}
\begin{center}
\begin{tabular}{cccc}
\tableline\tableline
Cluster & Date Observed & Exposure Times & FWHM \\
\tableline
NGC 2437 & October 31, 2010 & 2 sec + 95 sec $(u*)$ & 1.0 arcsec \\
 & & 1 sec + 50 sec $(g')$ & 0.9 arcsec \\
 & & 1 sec + 60 sec $(r')$ & 0.9 arcsec \\
\tableline
\end{tabular}
\end{center}
\caption{Log of Observations}
\end{table*}

\clearpage

\begin{figure}
\figurenum{1}
\epsscale{0.75}
\plotone{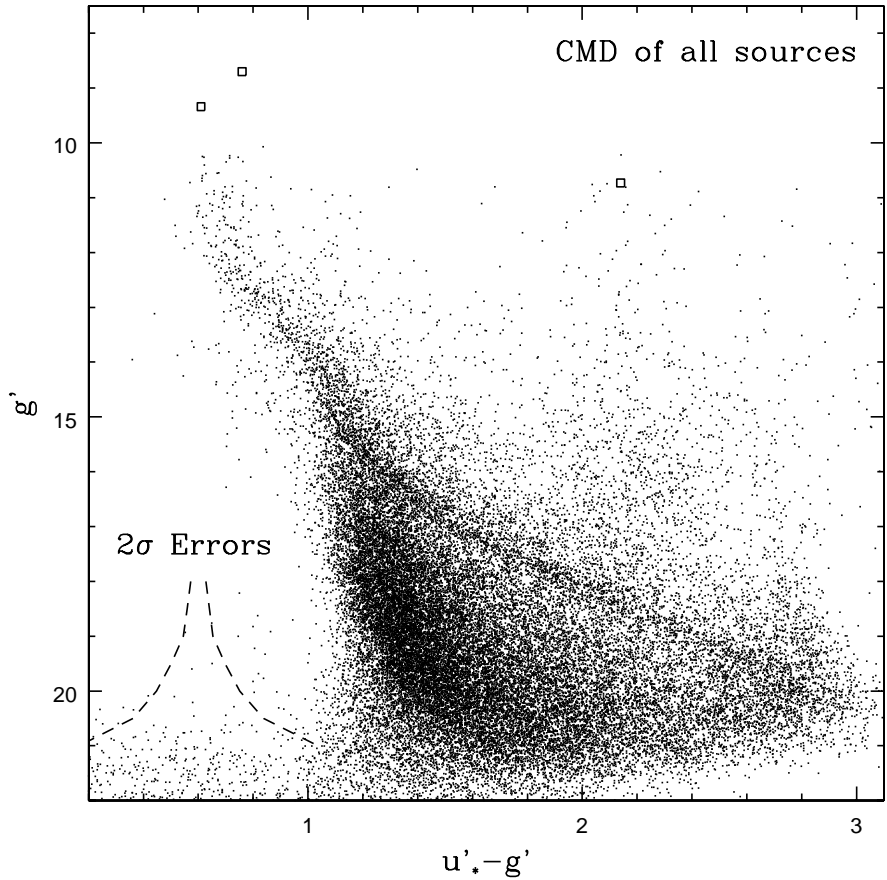}
\caption{The $(g', u'_*-g')$ CMD of objects in the MegaCam science field. The 
envelope of $\pm 2\sigma$ random errors calculated 
from the artificial star experiments is indicated. 
The open squares are based on the photoelectric $UBV$ measurements of 
stars with $V < 10.5$ from Lynga (1960), which were transformed into the SDSS 
system using relations from Smith et al. (2002) and Jordi et al. (2006).
While not the most prominent feature in the CMD, the main sequence of NGC 2437 runs 
diagonally across the CMD, and can be traced to $g' \sim 20$. The pronounced 
structure with $u'_*-g' \sim 1.2 - 1.6$ and $g' > 16$
is populated by main sequence field stars with blue colors.}
\end{figure}

\clearpage

\begin{figure}
\figurenum{2}
\epsscale{0.75}
\plotone{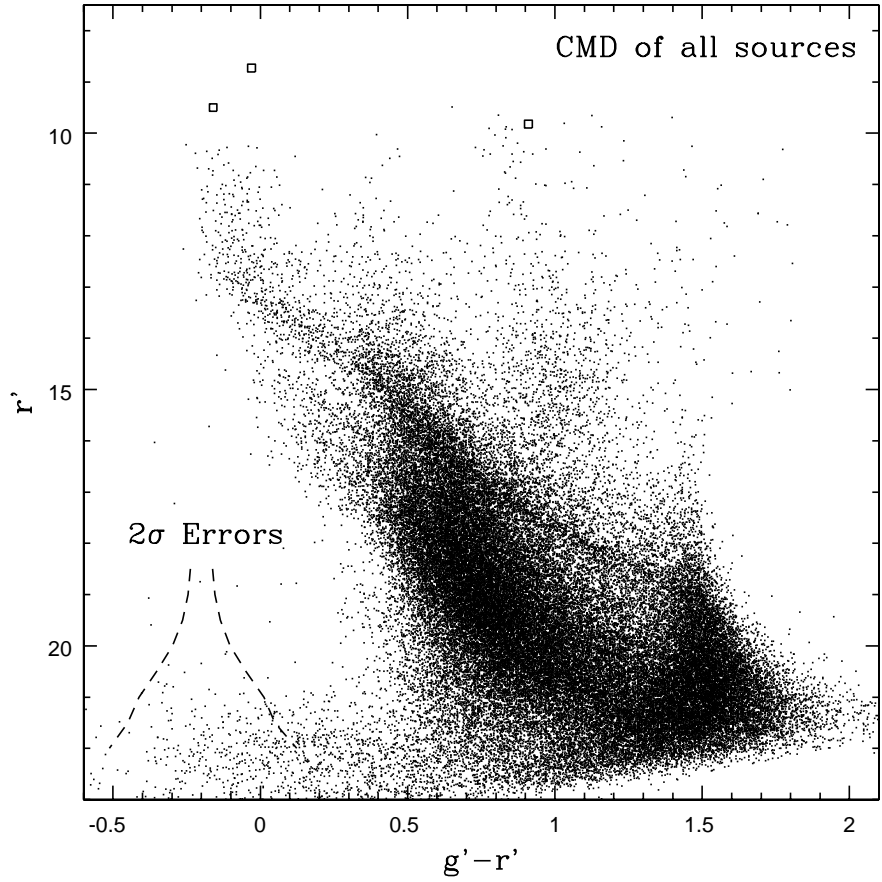}
\caption{The same as Figure 1, but showing the $(r', g'-r')$ CMD. 
The plume of sources in the lower right hand corner of the CMD with $g'-r' = 
1.5$ is populated by low mass main sequence stars in the field.}
\end{figure}

\clearpage

\begin{figure}
\figurenum{3}
\epsscale{0.75}
\plotone{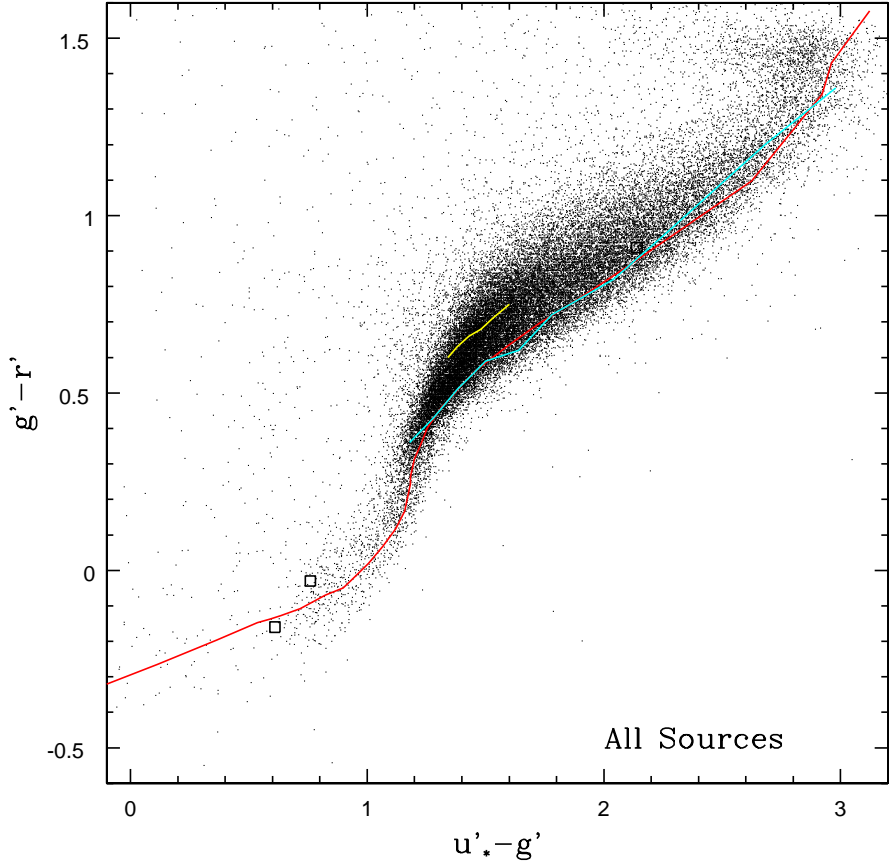}
\caption{The $(u'_*-g',g'-r')$ two color diagram (TCD). $UBV$ measurements from 
Lynga (1960), transformed into the SDSS system using relations from Smith 
et al. (2002) and Jordi et al. (2006), are plotted as open squares. 
The cyan line is the NGC 2437 main sequence traced from the CMDs, while 
the red line is the locus of solar neighborhood stars from Johnson (1966), 
reddened to match the mean $E(B-V)$ of NGC 2437 and transformed 
into the SDSS system using relations from Smith et al. (2002) and Jordi et al. 
(2006). The agreement between these two sequences suggests that the $u'_*$ 
magnitudes track $u'$ magnitudes over most of the color range. The yellow 
line is the ridgeline of stars in the field star plume. 
Objects with the reddest $g'-r'$ colors at a given $u'_*-g'$ tend to be 
non-cluster objects.}
\end{figure}

\clearpage

\begin{figure}
\figurenum{4}
\epsscale{0.75}
\plotone{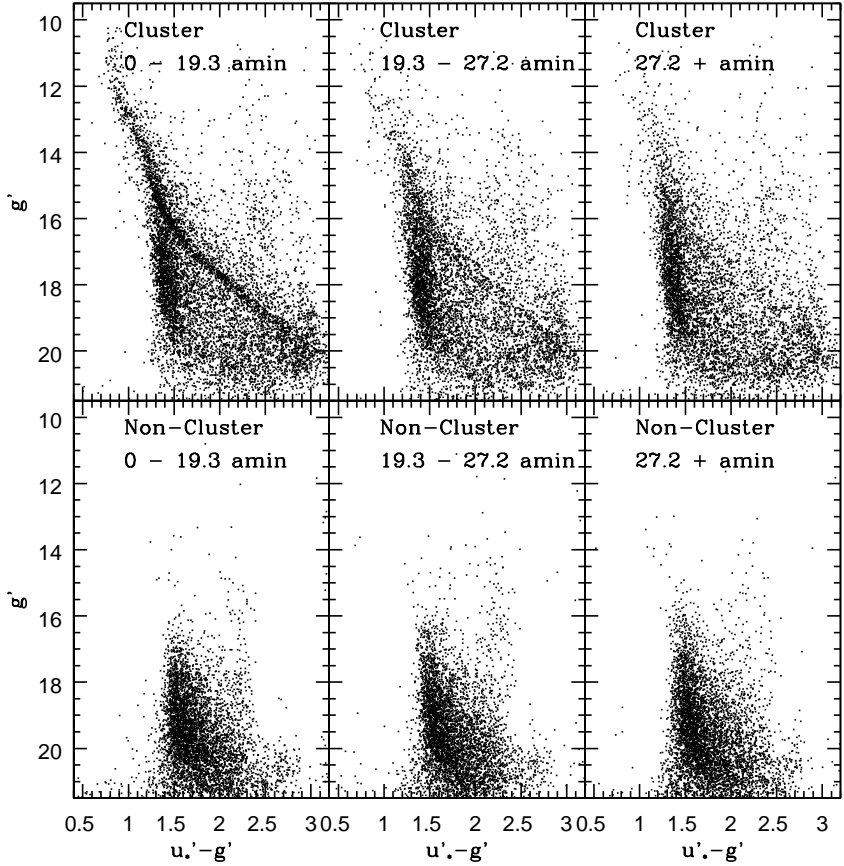}
\caption{(Top Row) The $(g', u'_*-g')$ CMDs of NGC 2437 cluster members, selected 
according to location on the $(u'_*-g', g'-r')$ TCD. The three radial intervals 
sample equal areas on the sky. The main sequence of NGC 2437 is clearly evident in 
the inner and middle annulus data, but is poorly defined in the outer annulus CMD. 
The bluest sources in the field star plume have been identified erroneously as 
`cluster' objects as their locations on the TCD fall within the extraction 
envelope for cluster stars. (Lower row) The CMDs of objects with SEDs at 
visible wavelengths that depart from those of cluster main sequence stars. 
The majority of non-cluster objects are faint and red. There is a more-or-less constant 
number of non-cluster objects in each radial interval, indicating that these objects 
are distributed uniformly across the field.}
\end{figure}

\clearpage

\begin{figure}
\figurenum{5}
\epsscale{0.75}
\plotone{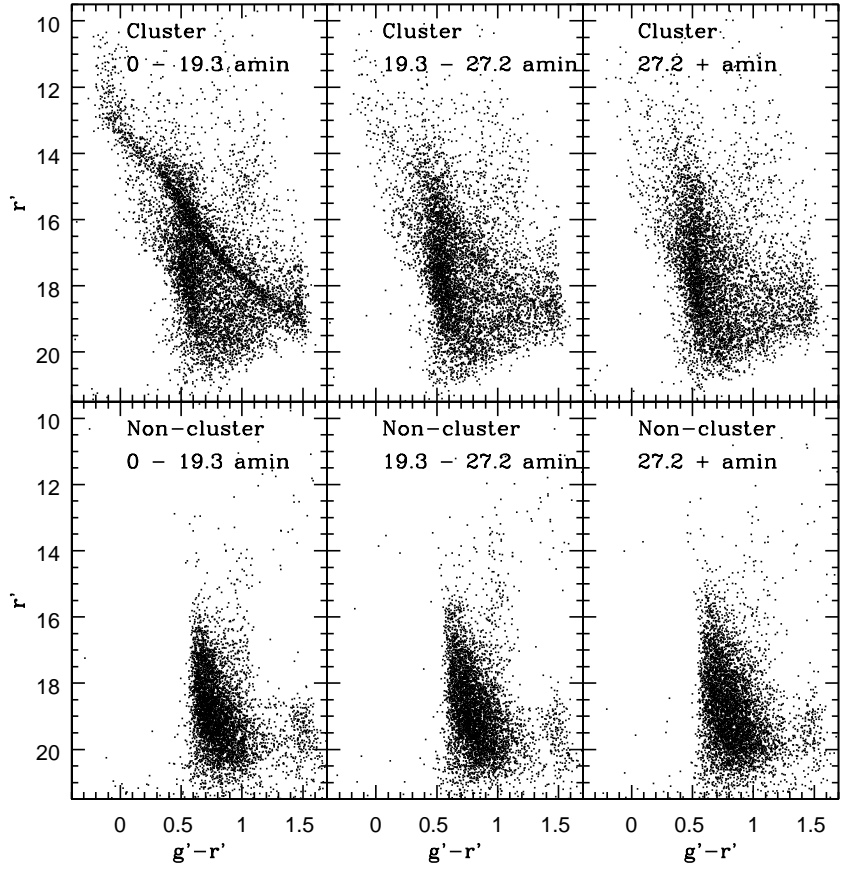}
\caption{The same as Figure 4, but showing the $(r',g'-r')$ CMDs.}
\end{figure}

\clearpage

\begin{figure}
\figurenum{6}
\epsscale{0.75}
\plotone{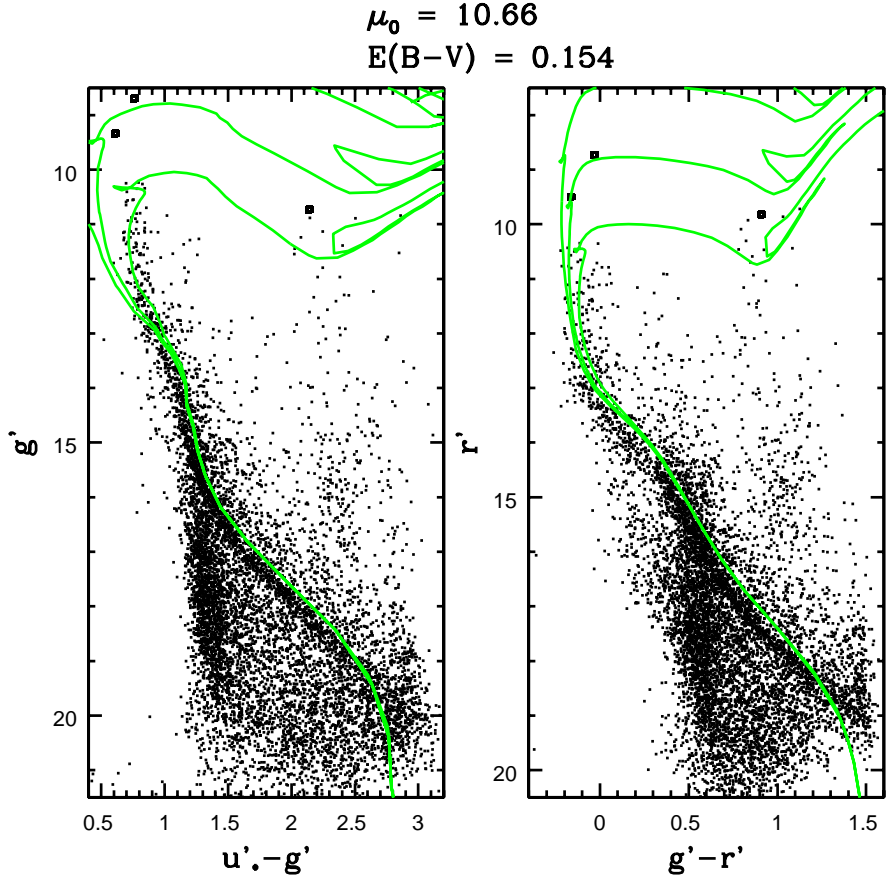}
\caption{The CMDs of cluster members in the inner annulus of NGC 2437 
are compared with Z = 0.019 isochrones from Marigo et al. (2008). 
Sequences with ages 80, 160, and 320 Myr are shown. Isochrone $u'$ magnitudes 
have been transformed into $u'_*$ magnitudes using the procedure described in 
the text. The distance modulus and reddening adopted 
for this comparison -- which are those in the WEDBA database as of mid-2012 -- 
are shown at the top of the figure. The models pass well above the main sequence 
at intermediate magnitudes, indicating that the apparent distance modulus is 
too small.}
\end{figure}

\clearpage

\begin{figure}
\figurenum{7}
\epsscale{0.75}
\plotone{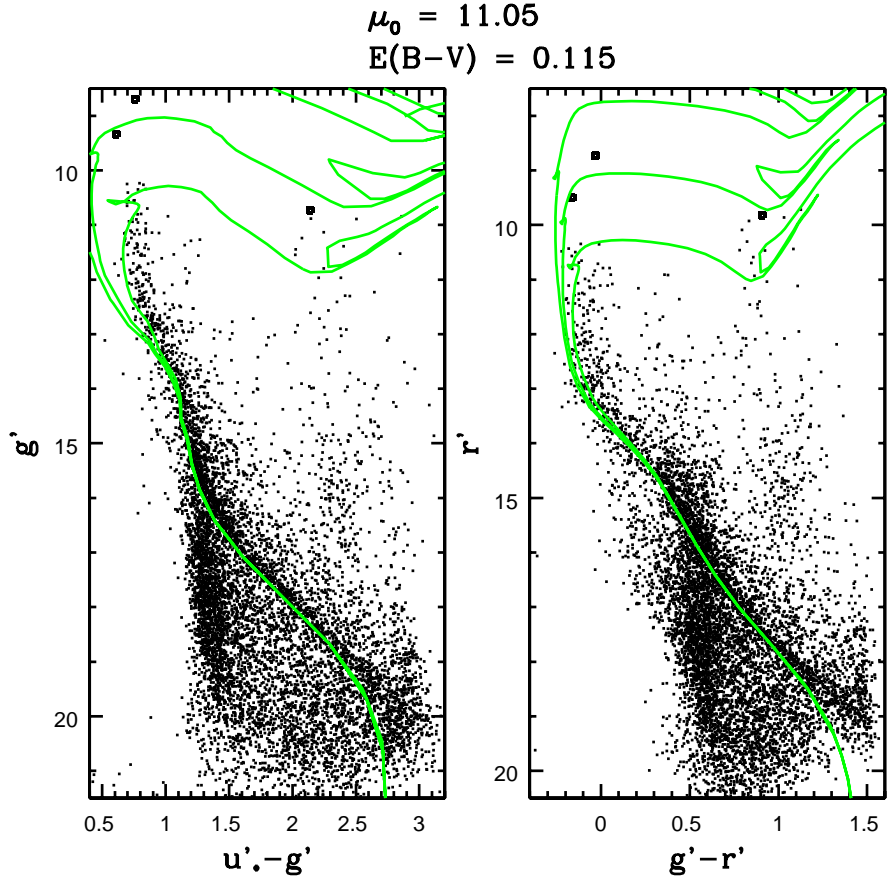}
\caption{The same as Figure 6, but assuming an absolute distance modulus 
$\mu_0 = 11.05$ (i.e. $r = 1.62$ kpc) and $E(B-V) = 0.115$. Note the improved 
agreement between the models and observations when compared with Figure 6. 
The models are consistent with an age log(t$_{yr}) \sim 8.35 \pm 0.10$.}
\end{figure}

\clearpage

\begin{figure}
\figurenum{8}
\epsscale{0.75}
\plotone{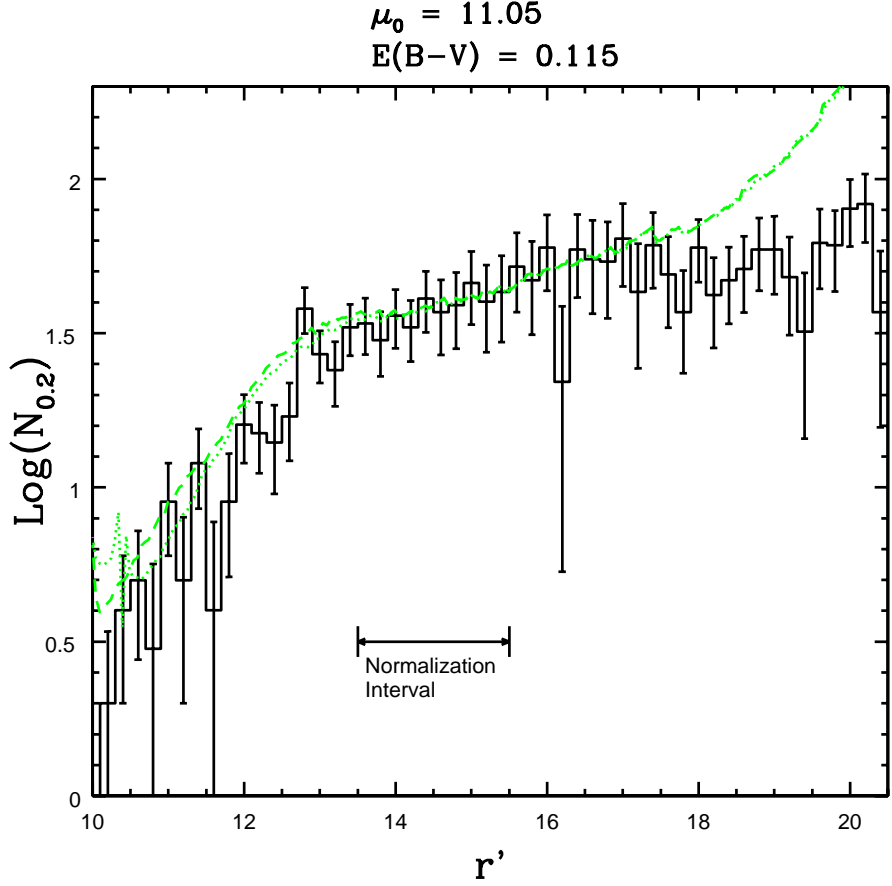}
\caption{The $r'$ luminosity function (LF) of main sequence stars within 19.3 arcmin of 
the cluster center. $N_{0.2}$ is the number of main sequence stars per 0.2 magnitude 
interval. The LF has been corrected for non-cluster objects by subtracting number 
counts measured at distances in excess of 27 arcmin 
from the cluster center, where the density of cluster stars is modest (Figures 
4 and 5). The error bars show uncertainties based on 
counting statistics. The green lines are solar metallicity models 
with a Kroupa (1998) IMF and ages 200 Myr (dashed line) and 
300 Myr (dotted line). The models have been shifted 
vertically to match the cluster LF between $r' = 13.5$ and 15.5. 
The models (1) match the change in the slope of the LF near $r' \sim 13$, which is due 
to an inflexion in the mass--magnitude relation and provides a loose check on the adopted 
distance modulus, and (2) clearly exceed the observed star counts when $r' > 17$.}
\end{figure}

\clearpage

\begin{figure}
\figurenum{9}
\epsscale{0.75}
\plotone{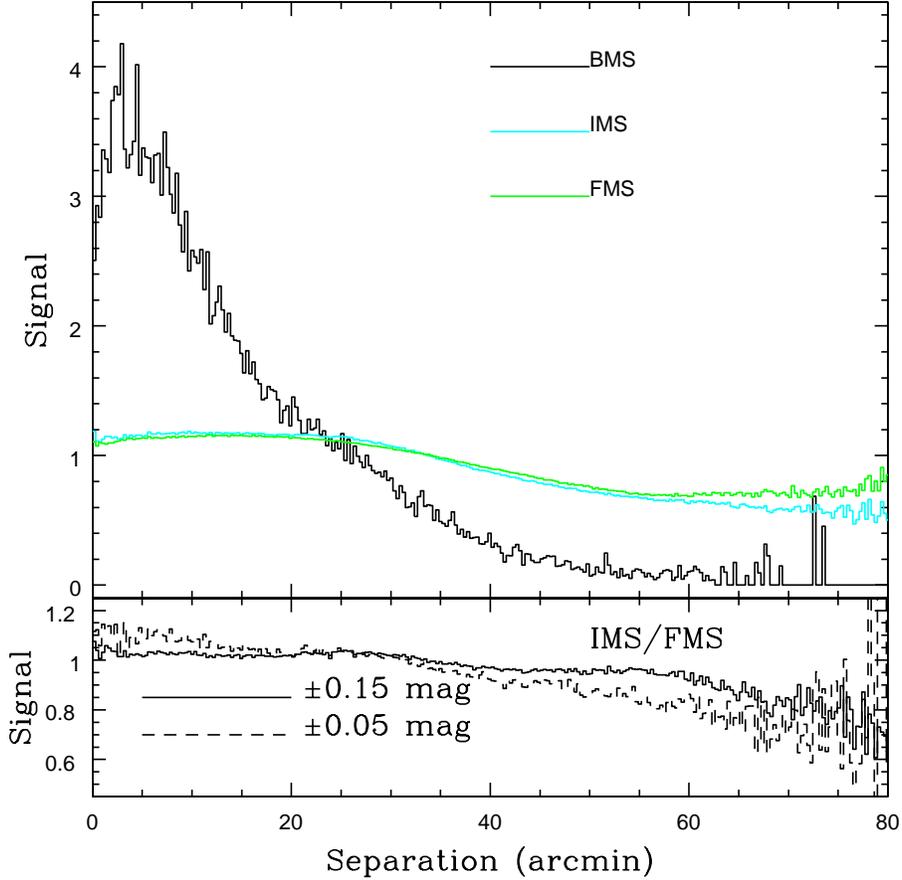}
\caption{(Top panel) The TPCFs of main sequence stars in 
NGC 2437, selected according to location on the $(r', g'-r')$ CMD. The TPCFs were 
constructed by dividing the separation functions (SFs) 
of objects in the BMS, IMS, and FMS samples by the SF of 
the FS sample, and then scaling the results according to the number of pairings. 
The TPCF of the BMS sample has by far the largest amplitude, and 
there is an inflexion point in the TPCF of the BMS sample 
at 20 -- 25 arcmin separations, suggesting a change in the structure of NGC 2437 at 
this point. (Lower panel) The ratio of the IMS and FMS SFs. 
Ratios are shown for $\pm 0.15$ mag (solid line) and 
$\pm 0.05$ mag (dashed line) extraction windows along the $(g'-r')$ axis, so that the 
effect of contamination from non-cluster sources on the TPCFs of these samples can be 
assessed. That the two ratios do not agree suggests that the relative 
distributions of the IMS and FMS samples are affected by contamination from non-cluster 
sources. The objects in the IMS and FMS samples obtained with the narrower 
CMD extraction window -- which should have a lower level of non-cluster contamination -- 
have different clustering properties, in a manner that 
is consistent with objects in the IMS sample being more centrally concentrated 
than those in the FMS sample.}
\end{figure}

\clearpage

\begin{figure}
\figurenum{10}
\epsscale{0.75}
\plotone{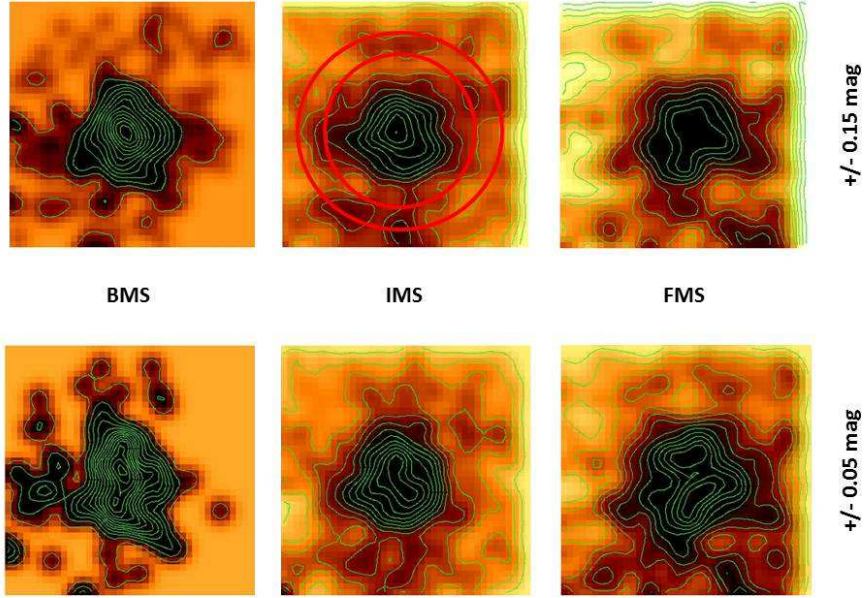}
\caption{The on-sky distribution of objects in the BMS, IMS, and FMS samples. 
The distributions of sources that fall within a $\pm 0.15$ mag extraction window 
about the main sequence locus on the $(r', g'-r')$ CMD are shown in the 
top row, while the distributions of objects within a 
$\pm 0.05$ mag extraction window are shown in the bottom row. 
The pixel intensities reflect the numbers of sources in $1.5 \times 1.5$ arcmin$^2$ 
gathers. North is at the top, and East is to the left. The red circles in the middle 
panel of the upper row show the radial boundaries used in the analysis of the CMDs and 
the LFs. The spatial distributions of sources obtained with the two extraction windows 
are in broad agreement, suggesting that structural features are due to cluster 
stars, rather than field stars. The distribution of sources in the BMS sample outside 
of the main body of the cluster is dominated by single objects. 
The FMS sample is more diffusely distributed than the IMS 
sample, and has a less regular morphology. FMS sources extend 
to the south and west of the cluster, and in \S 7 it is suggested that these might 
be objects that have been stripped from the cluster. The PN NGC 2438 is located 
slightly above and to the left of the MegaCam field center, and there is a localized 
concentration of objects near the PN in the FMS sample that is obtained with the 
$\pm 0.05$ mag extraction window.}
\end{figure}

\clearpage

\begin{figure}
\figurenum{A1}
\epsscale{0.75}
\plotone{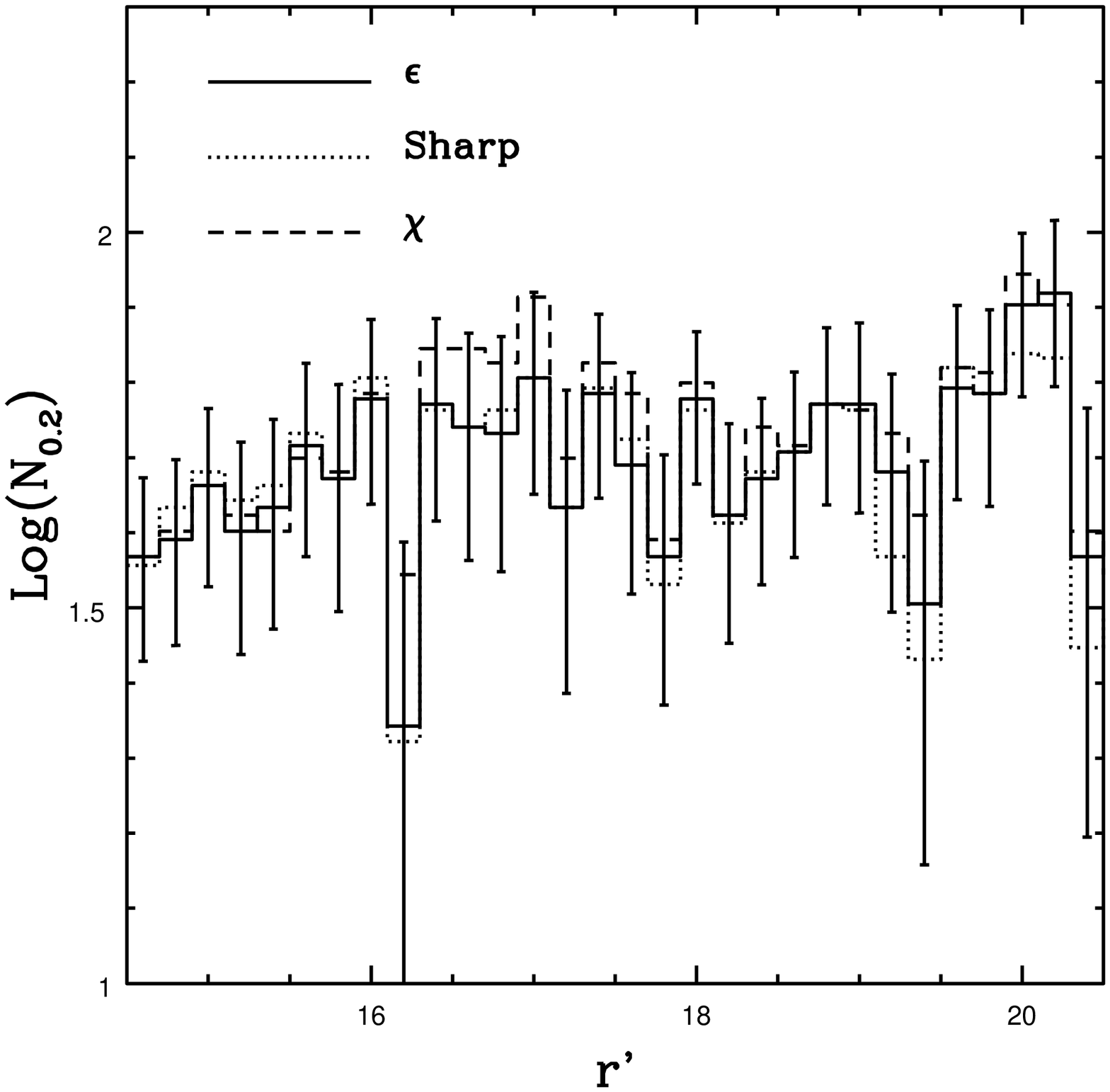}
\caption{The $r'$ LFs of main sequence stars in NGC 2437, obtained by using 
different indices computed by ALLSTAR to identify objects with suspect photometry. 
The error bars show uncertainties based on counting statistics 
computed in the data culled using $\epsilon$. The LFs have been 
corrected for non-cluster objects using the procedure described in 
\S 5. Sources have been excised that depart from 
the dominant trends with magnitude as defined by: $\epsilon$ 
(solid line), sharp (dotted line) and $\chi$ (dashed line). The LF-to-LF dispersion 
falls within the uncertainties computed for the baseline dataset discussed 
in \S 5, in which sources are culled using $\epsilon$. The flat nature of the 
NGC 2437 LF near the faint end is thus not due to an obvious bias introduced by the 
procedure used to remove objects with poor photometry.}
\end{figure}


\clearpage

\begin{references}

\reference{}Balona, L. A., \& Lacey, C. D. 1996, MNRAS, 281, 1341

\reference{}Bastian, N., \& Goodwin, S. P. 2006, MNRAS, 369, L9

\reference{}Bastian, N., Weisz, D. R., Skillman, E. D., et al. 2011, MNRAS, 412, 1539

\reference{}Bonatto, C., \& Bica, E. 2011, MNRAS, 415, 2827

\reference{}Boulade, O., et al. 2003, Proc. SPIE, 4841, 72

\reference{}Chumak, Y. O., \& Rastorguev, A. S. 2006, Astron. Lett., 32, 446

\reference{}Chumak, Y. O., Platais, I., McLaughlin, D. E., Rastorguev, A. S., \& Chumak, O. V. 2009, MNRAS, 402, 1841

\reference{}Clem, J. L., VandenBerg, D. A., \& Stetson, P. B. 2008, AJ, 682

\reference{}Dahm, S. E. 2005, AJ, 130, 611

\reference{}Davidge, T. J. 2010, ApJ, 725, 1342

\reference{}Davidge, T. J. 2012, ApJ, 761, 155

\reference{}Davidge, T. J., \& Puzia, T. H. 2011, ApJ, 738, 144

\reference{}Davidge, T. J., Puzia, T. H., \& McConnachie, A. 2011, ApJ, 728, L23

\reference{}Davidge, T. J., McConnachie, A. W., Fardal, M. A. et al. 2012, ApJ, 751, 74
 
\reference{}Delgado, A. J., Gonzalez-Martin, O., Alfaro, E. J., \& Yun, J. 2006, ApJ, 646, 269

\reference{}Delgado, A. J., Alfaro, E. J., \& Yun, J. L. 2011, A\&A, 531, A141

\reference{}Dworetsky, M. M. 1975, AJ, 80, 131

\reference{}Harris, G. L. H., Fitzgerald, M. P. V., Mehta, S., \& Reed, B. C. 1993, AJ, 106, 1533

\reference{}Ianna, P. A., Adler, D. S., Faudree, E. F. 1987, AJ, 92, 347

\reference{}Johnson, H. L. 1966, ARA\&A, 4, 193

\reference{}Jordi, K., Grebel, E. K., \& Ammon, K. 2006, A\&A, 460, 339

\reference{}Kholopov, P. N. 1969, Sov. Astron., 12, 625

\reference{}Kiss, L. L., Szabo, G. M., Balog, Z., Parker, Q. A., \& Frew, D. J. 2008, MNRAS, 391, 399

\reference{}Kroupa, P. 1998, in Brown Dwarfs and Extrasolar Planets, ASP Conf. 134. 483

\reference{}Lada, C. J., \& Lada, E. A. 2003, ARAA, 41, 57

\reference{}Lamers, H. J. G. L. M., Gieles, M., Bastian, N., Baumgardt, H., Kharchenko, N. V., \& Portegies Zwart, S. 2005, A\&A, 441, 117

\reference{}Larson, R. B. 1995, MNRAS, 272, 213

\reference{}Lynga, G. 1960, ArA, 2, 379

\reference{}Marigo, P., Girardi, L., Bressan, A. et al. 2008, A\&A, 482, 883

\reference{}Mermilliod, J.-C. 1995, ASSL, 203, 127

\reference{}Moitinho, A., Alves, J., Huelamo, N., \& Lada, C. J. 2001, ApJ, 563, L73

\reference{}Perry, C. L. 1973, in Spectral Classification and Multicolor Photometry, IAUS, 50, 192

\reference{}Prisinzano, L., Mirela, G., Sciortino, S., \& Favata, F. 2003, A\&A, 404, 927

\reference{}Rojo Arellano, E., Pena, J. H., Gonzalez, D. 1997, A\&AS, 123, 25

\reference{}Sagar, R., \& Richtler, T. 1991, A\&A, 250, 324

\reference{}Salpeter, E. E. 1955, ApJ, 121, 161

\reference{}Schlegel, D. J., Finkbeiner, D. P., \& Davis, M. 1998, ApJ, 500, 525
 
\reference{}Sharma, S., Pandey, A. K., Ogura, K., Mito, H., Tarusawa, K., \& Sasov, R. 2006, AJ, 132, 1669

\reference{}Sharma, S., Pandey, A. K., Ogura, K., Aoki, T., Pandey, K., Sandhu, T. S., \& Sagar, R. 2008, AJ, 135, 1934

\reference{}Shobbrook, R. R. 1984, MNRAS, 211, 659

\reference{}Simon, M. 1997, ApJ, 482, L81

\reference{}Smith, J. A., Tucker, D. L., Kent, S., et al. 2002, AJ, 123, 2121

\reference{}Smith, J., Fellhauer, M., Goodwin, S., \& Assmann, P. 2011, MNRAS, 414, 3036

\reference{}Stetson, P. B. 1981, AJ, 86, 1500

\reference{}Stetson, P. B. 1987, PASP, 99, 191

\reference{}Stetson, P. B., \& Harris, W. E. 1988, AJ, 96, 909

\reference{}Stetson, P. B., Bruntt, H., \& Grundahl, F. 2003, PASP, 115, 413

\reference{}Tognelli, E., Prada Moroni, P. G., \& degl'Innocenti, S. 2011, A\&A, 533, A109

\end{references}
\end{document}